# Second Order Catalytic Quasispecies Yields First Order Phase Transition

Nathaniel Wagner, Emmanuel Tannenbaum* and Gonen Ashkenasy*

*Department of Chemistry, Ben Gurion University of the Negev, Beer Sheva 84105, Israel*

**Abstract**

The quasispecies model describes processes related to the origin of life and viral evolutionary dynamics. We discuss how the error catastrophe that reflects the transition from localized to delocalized quasispecies population is affected by catalytic replication of different reaction orders. Specifically, we find that $2^{nd}$ order mechanisms lead to $1^{st}$ order discontinuous phase transitions in the viable population fraction, and conclude that the "higher" the interaction the "lower" the transition. We discuss potential implications for understanding the replication of highly mutating RNA viruses.

*Corresponding authors
Emails: emanuelt@bgu.ac.il (ET), gonenash@bgu.ac.il (GA)

The quasispecies model can be formulated as a set of differential equations describing the evolution of self-replicating molecules.[1-2] These equations were originally developed by Eigen and Schuster [3-4] as a way to model molecular evolutionary processes related to the origin of life. In addition to molecular evolution, the quasispecies model has also been extensively applied toward the understanding of viral evolutionary dynamics, in particular that of rapidly mutating single-stranded RNA viruses (e.g. HIV).[5-7]

Likewise, catalytic reactions have been shown to be relevant to the origin of life and early molecular evolution by facilitating self-replication and dynamic evolution.[8-12] Such reactions can operate both autocatalytically and cross catalytically, where autocatalysis is a mechanism for self-replication and cross catalysis may lead to mutation [12]. It has been further shown how interacting catalytic reactions of various reaction order can form catalytic networks of increasing complexity.[13-15] In a recent paper [16] we have shown how certain crucial features of network complexity require at least second order catalysis [3,12,17]. With the goal of developing a more complete understanding of evolutionary dynamics in higher order autocatalytic networks, we have developed a quasispecies approach for analyzing mutation and selection in catalytic reactions of varying order.

Typically, quasispecies models are characterized by an upper mutational threshold, beyond which natural selection can no longer localize the population about the fittest sequences.[18-20] Below this error threshold, the population consists of a cloud of related strains termed a quasispecies, while above this error threshold the evolutionary dynamics is characterized by random genetic drift over the sequence space. The transition from a localized population distribution to a delocalized distribution is known as the error catastrophe.

In this Letter we show how the error catastrophe is affected when different replication models are considered. To do so, we determine how models of replication dynamics characterized by a catalyst and a template species influence the nature of the transition at the error threshold. Most interestingly, we find that $2^{nd}$ order catalytic mechanisms lead to $1^{st}$ order phase transitions with respect to the viable population fraction. Understanding the mutation-selection balance and the error threshold in catalytic networks of varying order is both theoretically interesting, and has potentially important implications in understanding cooperative phenomena in the replication mechanisms of highly mutating genetic sequences, such as RNA viruses.

We therefore propose a generalized two-stage cooperative catalytic model (Eq. 1), which assumes a population of single-stranded sequences of some biopolymer (e.g. RNA or protein), denoted by $\sigma_i$, where each $\sigma_i$ represents a distinct sequence that can both catalyze the replication of other sequences, as well as act as a template for the production of new sequences:

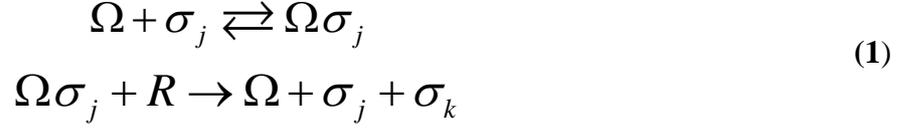

$$\Omega + \sigma_j \rightleftarrows \Omega\sigma_j$$
$$\Omega\sigma_j + R \to \Omega + \sigma_j + \sigma_k$$
(1)

By using this general model we consider three separate cases, characterized by the following replication schemes: (i) $\Omega = M$, describing the obligatory binding of $\sigma_j$ to a non-genomic molecule (e.g. small molecule) prior to replication, (ii) $\Omega = \sigma_j$, reflecting the formation of a homo-dimer catalyst for replication, and (iii) $\Omega = \sigma_i$ for formation of hetero-dimer catalyst for replication. In the following we first formulate the rate laws for each of the cases, and then solve them analytically for the single fitness peak landscape approximation. Since we note that case (ii) can be viewed as a specific example of case (iii), we discuss both of them together, and highlight the differences where appropriate.

*Case (i)* - The first step consists of a reversible binding reaction $M + \sigma_j \leftrightarrow M\sigma_j$, characterized by a forward first order kinetics rate constant $b_j$, and a dissociation rate constant $d_j$. In the second step, the complex $M\sigma_j$ catalyzes the formation of a daughter sequence $\sigma_k$ from environment raw materials, $R$, with rate constant $c_{jk}$. Since $\sigma_j$ acts as a catalyst for its replication, we assume that $c_{jk}$ depends on a fitness factor $f_j$ associated with sequence $\sigma_j$. Furthermore, we assume that the replication of $\sigma_j$ is not necessarily error-free, so there exists a probability $p_{jk}$ that the daughter of $\sigma_j$ will be $\sigma_k$. When $j = k$ the sequence $\sigma_j$ has self-replicated, and when $j \neq k$ a new sequence $\sigma_k$ has been formed by mutation. In general, $p_{jk}$ will decrease with the Hamming distance $D_H(\sigma_j, \sigma_k)$ between the two sequences, which is the number of mutations by which $\sigma_j$ differs from $\sigma_k$.[2]

In the steady-state limit $d/dt\,[M\sigma_j] = 0$ corresponding to Michaelis-Menten kinetics with low intermediate concentration, or alternatively, in the fast equilibrium limit while assuming $f_j p_{jk}[R] \ll d_j$, the mechanism of Eq. (1) corresponds to the simpler one-step model, where the $f_j$ are rescaled:

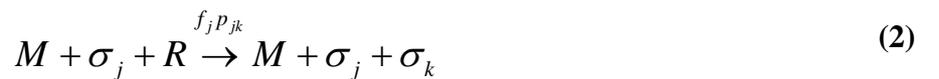

$$M + \sigma_j + R \xrightarrow{f_j p_{jk}} M + \sigma_j + \sigma_k$$
(2)

If $f_j p_{jk}[R] \gg d_j$, corresponding to a very high-resource environment [21] with a large fitness factor and high mutation rate, the rate constant will simply be $b_j$ and the kinetics will be independent of

[R]. In all cases the reaction is first order in M, whether [M] is high or low. The quasispecies equations for the system described by Eq. (2) are:

$$\frac{1}{[\sigma]}\frac{dx_l}{dt} = \sum_j f_j p_{jl} x_j - x_l \sum_j f_j x_j \qquad (3)$$

where $[\sigma] \equiv \sum_k [\sigma_k]$, and $x_l = [\sigma_l]/[\sigma]$ gives the relative concentration of a given sequence $\sigma_l$. A similar model has previously been solved and shown to yield continuous phase transitions with respect to the viable population fraction with varying mutation rate[2,22].

*Cases (ii) and (iii)* - In these cases the second order catalysis may use either homo-dimer ($i = j$) or hetero-dimer ($i \neq j$) intermediate complexes. The first step consists of the reversible reaction $\sigma_i + \sigma_j \leftrightarrow \sigma_i\sigma_j$, for which the forward reaction follows second order kinetics with binding rate constant $b_{ij}$, and the reverse reaction has dissociation rate constant $d_{ij}$. Since the binding is mainly the result of diffusion processes, we can assume $b_{ij}$ to be independent of $i$ and $j$, so $b_{ij} = b$. On the other hand, $d_{ij}$ is dependent on the "fit" between the two sequences, and in general, $d_{ij}$ will increase with sequence dissimilarity of $\sigma_i$ and $\sigma_j$. In the second step, $\sigma_i$ catalyzes the formation of a daughter sequence $\sigma_k$ from R, using $\sigma_j$ as a template. Note that for these cases we distinguish between catalyst and template, and thus the complex $\sigma_i\sigma_j$ is not equivalent to $\sigma_j\sigma_i$. Since $\sigma_i$ acts as the catalyst, we assume that $c_{ijk}$ depends on a fitness factor $f_i$ associated with sequence $\sigma_i$. Furthermore, as for case (i), the replication of $\sigma_j$ is not necessarily error-free, so we use a probability $p_{jk}$ that the daughter of $\sigma_j$ will be $\sigma_k$, and accordingly, $c_{ijk} = f_i p_{jk}$. Here again, both the steady-state limit $d/dt [\sigma_i\sigma_j] = 0$, or the fast equilibrium limit assuming $f_i p_{jk}[R] \ll d_{ij}$, lead to a simpler one-step model[23], expressed using an "affinity" term $a_{ij} = b/(d_{ij} + f_i)$:

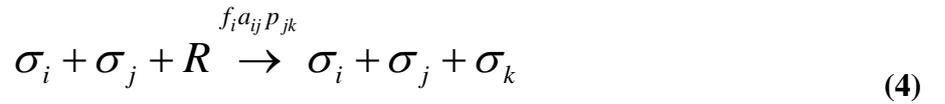

$$\sigma_i + \sigma_j + R \xrightarrow{f_i a_{ij} p_{jk}} \sigma_i + \sigma_j + \sigma_k \qquad (4)$$

For $f_i p_{jk} [R] \gg d_{ij}$, corresponding to a very high-resource environment [21] with a large fitness factor and high mutation rate, the rate constant will be $b_{ij}$ and the kinetics independent of [R]. The quasispecies equations for the Eq. (4) system are:

$$\frac{1}{[\sigma]}\frac{dx_l}{dt} = \sum_i f_i x_i \sum_j a_{ij} p_{jl} x_j - x_l \sum_i f_i x_i \sum_j a_{ij} x_j \qquad (5)$$

*Analytical solutions* - The three above cases can be analytically solved for the single fitness peak landscape approximation [2,22], yielding a tractable closed-form solution for a simple landscape, as well as qualitative insight into the general solutions. Such a landscape assumes a single viable master sequence. In order to compute its concentration, we use the following approximation: we divide the sequences into two classes, where $l = 0$ corresponds to the master sequence and $l = 1$ corresponds to all other sequences. Since the master sequence is the viable one, $f_0 > 1$ while $f_1 = 1$. We further assume, for the second order cases (ii and iii), a larger affinity between identical sequences, so $a_{00} > a_{01} = a_{10} \approx a_{11}$ (since most $l = 1$ cases correspond to different sequences), and define the ratio $a = a_{00} / a_{01} > 1$. Furthermore, we use a probability $1 - p$ for the mutation of the master sequence and assume an infinitesimal probability of back mutation, i.e., $p_{00} = p$ and $p_{01} = 1 - p$, while $p_{10} = 0$ and $p_{11} = 1$.

For case (i), the $d/dt\,[x_l] = 0$ steady-state limit for Eq. (3) leads to the following two solutions, $x_0 = 0$ and $x_0 = (f_0\,p - 1) / (f_0 - 1)$. Since $x_0 > 1$, the zero solution will be relevant for $f_0\,p < 1$ and the second solution will be relevant for $f_0\,p > 1$. This is also consistent with the obvious constraints $x_0 = 0$ at $p = 0$ (all mutations) and $x_0 = 1$ at $p = 1$ (no mutations).

For cases (ii) and (iii), the $d/dt\,[x_l] = 0$ steady-state limit for Eq. (5) yields three solutions, $x_0 = 0$ and the quadratic equation:

$$\left[ f_0 (a-1) \right] x_0^2 + \left[ p(1 - f_0 a) + f_0 - 1 \right] x_0 + \left[ 1 - p \right] = 0 \tag{6}$$

In the vicinity of $p = 1$, $a = 1$ and $f_0 = 1$, using $\delta p = 1 - p$, $\delta a = a - 1$ and $\delta f_0 = f_0 - 1$, one obtains the first order approximation:

$$x_0 = \frac{1}{2} \pm \frac{1}{2}\sqrt{1 - 4\frac{\delta p}{\delta a}} \tag{7}$$

The zero solution will be relevant for small $p$, and the positive solution of the quadratic equation will be relevant for larger $p$, since only it corresponds to the $x_0 = 1$ at $p = 1$ constraint. Specifically for case (ii), since this corresponds to case (iii) for very large $a$, one can take this limit in Eq. (6) and obtain the solution $x_0 = p$.

A plot of the qualitative solutions for the three cases is shown in Figure 1. The solutions clearly show the phase transitions, where the zero solution switches to the other relevant solution. While for case (i) the transition is a continuous one (as expected), and for case (ii) there is actually no transition at all, case (iii) displays a discontinuous first order phase transition. By solving Eq. (6) numerically, quantitative solutions for case (iii) were computed. A few representative solutions, for several combinations of $a$ and $f_0$, are plotted in Figure 2, clearly showing the discontinuous phase transition and confirming the above analysis. Note that for large $a$ the results approach those of case (ii).

Insert Figure 1 here

Insert Figure 2 here

In order to rigorously prove the existence of a phase transition in case (iii), it is necessary to show that the steady-state solution for the localization length diverges at the critical values of mutation rate. For this purpose we write the transition probability $p_{jk}$ from sequence $j$ to sequence $k$ as follows:

$$p_{jk} = \left(\frac{\varepsilon}{S-1}\right)^{D_H(\sigma_j,\sigma_k)} (1-\varepsilon)^{L-D_H(\sigma_j,\sigma_k)} \tag{8}$$

where $\varepsilon$ is the probability that a given base will mutate, $L$ is the length of each sequence (i.e., number of bases), $S$ is the alphabet size (i.e, number of possible bases), while recalling that altogether there are $N = S^L$ possible sequences. When studying steady-state behavior for long sequence lengths, it is appropriate to consider the behavior in the $L \to \infty$ limit. Defining $\mu = \varepsilon L$, where $\mu$ is the average total number of replication errors per replication cycle, while holding $\mu$ constant in the $L \to \infty$ limit, the probability of correctly replicating a sequence is given by $(1-\varepsilon)^L \to e^{-\mu}$. Similarly, for finite values of $l$, $(1-\varepsilon)^{L-l} \to e^{-\mu}$.

It is convenient here to classify the relative concentrations by Hamming class, where the Hamming class $C_H(l)$ consists of all sequences at a given Hamming distance $l$ from the master sequence. Each Hamming class $C_H(l)$ contains $C_l$ distinct sequences, and the relative concentrations classified by Hamming class are given by:

$$z_l = \sum_{\sigma_i \in C_H(l)} x_i = C_l x_i = \binom{L}{l}(S-1)^l x_i \tag{9}$$

since by symmetry all $x_i$ within the same Hamming class are equal (assuming $a_{ij}$ does not discriminate between distinct sequences of the same Hamming class).

The localization length [2] can now be computed:

$$\langle l \rangle = \sum_{l=0}^{L} l\, z_l \to \sum_{l=0}^{\infty} l\, z_l \tag{10}$$

where the sum is now over all Hamming distances, while taking the long sequence length limit. First we compute the total transition probability from Hamming class $j$ to Hamming class $l$. Since we neglect back mutations, exactly $l - j$ bases mutate and exactly $L - (l - j)$ bases do not mutate. Taking into account the number of ways of choosing $l - j$ mutating bases from among the $L - j$ bases not already mutated, and considering that each base can mutate in $S - 1$ ways, this yields:

$$P_{jl} = \binom{L-j}{l-j}(S-1)^{l-j}\left(\frac{\varepsilon}{S-1}\right)^{l-j}(1-\varepsilon)^{L-(l-j)} \xrightarrow{L\to\infty} \frac{\mu^{l-j}}{(l-j)!}e^{-\mu} \quad (11)$$

We can write the quasispecies equations (Eq.(5)) in terms of Hamming classes:

$$\frac{1}{[\sigma]}\frac{dz_l}{dt} = \sum_i f_i z_i \sum_j a_{ij} P_{jl} z_j - z_l \sum_i f_i z_i \sum_j a_{ij} z_j = \sum_i f_i z_i \sum_j a_{ij} \frac{\mu^{l-j}}{(l-j)!}e^{-\mu} z_j - \overline{f}(t) z_l \quad (12)$$

where the expression $\overline{f}(t) = \sum_i f_i z_i \sum_j a_{ij} z_j$ can be thought of as the mean fitness [2] as a function of time. As before, we approximate $f_i$ and $a_{ij}$ by assuming only two relevant values:

$$\overline{f}(t) = \sum_i \sum_j f_i a_{ij} x_i x_j = f_0 a x_0^2 + (f_0+1) x_0 (1-x_0) + (1-x_0)^2 \xrightarrow{x_0=0} 1 \quad (13)$$

Inserting the $l = 0$ and $l > 0$ cases into Eq. (12), differentiating and normalizing Eq. (10), and then taking the steady state limit $d/dt \langle l \rangle = 0$, yields:

$$\langle l \rangle = \mu \frac{f_0 x_0^2 (a-1) + (f_0-1) x_0 + 1}{\overline{f}(t) - (f_0-1) x_0 - 1} \quad (14)$$

Since $x_0 = 0 \Rightarrow \overline{f}(t) = 1$, we get $<l> \to \infty$ at the critical values of mutation rate. This rigorously proves the existence of a phase transition, which has already been shown to be discontinuous.

We have shown how a simple two-stage catalytic quasispecies model yields varying types of error catastrophes and phase transitions, depending on the nature and order of the catalytic reactions. By solving the second order hetero-dimer model, approximately and numerically, and by showing the divergence of the localization length, we have rigorously shown how this model yields a discontinuous first order phase transition with respect to the viable population fraction. Our results are consistent with several, significantly different works displaying discontinuous first order phase transitions in quasispecies models involving higher order interactions, such as (sexual) recombination [24-26], bulk [5,27-28] or infinite length interaction [29-30]. In contrast, finite first order models have displayed continuous phase transitions at steady state [1-2,4,24,28,31]. This leads us to empirically conclude that the "higher" the interaction the "lower" the transition, and vice-versa. This conclusion may at first seem surprising, even counterintuitive. However, many models display highly sensitive behavior with increasing cooperativity, and increasingly complex biochemical and metabolic networks have been shown to be more robust and versatile, allowing for sharper transitions. [32-33] Another representative example, far from the physical sciences, is the correlation between increasingly complex financial tools and mechanisms with volatility and the potential for economic catastrophe [34-35].

This paper's findings may be applied towards the replication and mutation of viruses, and in particular towards the evolution of RNA viruses whose mutation rates are very high. RNA is especially appropriate for our model, since it has been shown to function both as a template and as a catalyst. [36]. While there is no direct information regarding the population of viruses as a function of mutation rate, data is available on their survival under varying environmental conditions. One can then use this information to learn about the replicative mechanisms of the viruses. We suggest that a varying mutation rate in a constant environment is mathematically equivalent to a constant mutation rate in a changing environment (when these changes affect the relative survival of the species). Therefore, viruses showing a relatively flat response to varying environmental conditions would usually correspond to continuous transitions, suggesting simpler or lower order replicating mechanisms such as case (i) or even case (ii). Alternatively, viral populations that are very sensitive and perhaps unstable relative to the environment may correspond to discontinuous transitions, suggesting more complex or cooperative interactions and higher order mechanisms of replication such as in case (iii).

## Acknowledgments

We thank Zehavit Dadon for her assistance in writing the paper. This research was partially supported by a grant from the Israel Science Foundation (ISF 1291/08), and by the Edmond J Safra Center.
.

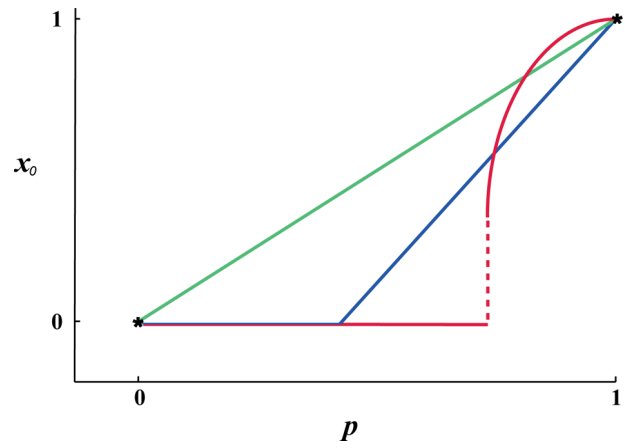

Figure 1. Qualitative solutions for the three cases of the two-stage catalytic quasispecies model. Case i (*blue*) shows a continuous phase transition, case ii (*green*) shows no phase transition, and case iii (*red*) yields a discontinuous 1$^{st}$ order phase transition.

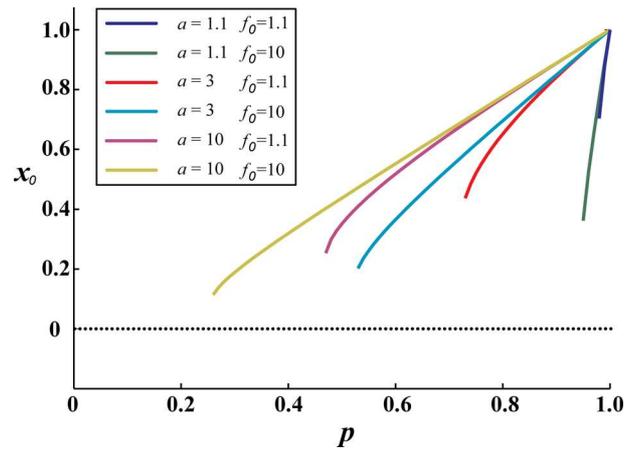

Figure 2. Numerical solutions of Equation (6) for several representative combinations of $a$ and $f_0$, clearly showing the discontinuous first order phase transition for case (iii).